%% file: main.tex
\documentclass[conference]{IEEEtran}
\IEEEoverridecommandlockouts

\IEEEoverridecommandlockouts

\usepackage{cite}

\usepackage{amsfonts}
\usepackage{amssymb}
\usepackage{algorithmic}
\usepackage{textcomp}
\usepackage{xcolor}
\def\BibTeX{{\rm B\kern-.05em{\sc i\kern-.025em b}\kern-.08em
    T\kern-.1667em\lower.7ex\hbox{E}\kern-.125emX}}

\usepackage{amsmath} 

\usepackage{textcomp}
\usepackage{xcolor}

\usepackage{graphics, epstopdf, graphicx, epsfig, psfrag, epsf,subfigure}

\usepackage{algorithm}
\usepackage{algorithmic}

\usepackage{url}

\newcommand{\ie}{{\em i.e., }}

\newcommand{\Nset}{\mathcal{N}}
\newcommand{\Hset}{\mathcal{H}}

\newcommand{\Lset}{\mathcal{L}}

\newcommand{\Mset}{\mathcal{M}}
\newcommand{\Tset}{\mathcal{T}}

\usepackage{xspace,exscale,relsize}

\newcommand{\ours}{{\tt{PA-MDI}}\xspace}

\begin{document} {



\pagestyle{empty}

\title{Priority-Aware Model-Distributed Inference \\ at Edge Networks}


\author{\IEEEauthorblockN{Teng Li}
\IEEEauthorblockA{\textit{Electrical and Computer Engineering} \\
\textit{University of Illinois Chicago}\\
tli81@uic.edu}
\and
\IEEEauthorblockN{Hulya Seferoglu}
\IEEEauthorblockA{\textit{Electrical and Computer Engineering} \\
\textit{University of Illinois Chicago}\\
hulya@uic.edu}
}
}



%
\maketitle
\input{abstract}
%
%

\input{intro}

\input{related}

\input{model}

\input{M2MDI}

\input{results}
\input{conclusion}
\bibliographystyle{IEEEtran}
\bibliography{refs}

\end{document}

%% file: abstract.tex
\begin{abstract}
Distributed inference techniques can be broadly classified into data-distributed and model-distributed schemes. In data-distributed inference (DDI), each worker carries the entire Machine Learning (ML) model but processes only a subset of the data. However, feeding the data to workers results in high communication costs, especially when the data is large. An emerging paradigm is model-distributed inference (MDI), where each worker carries only a subset of ML layers. In MDI, a source device that has data processes a few layers of ML model and sends the output to a neighboring device, i.e., offloads the rest of the layers. This process ends when all layers are processed in a distributed manner. In this paper, we investigate the design and development of MDI when multiple data sources co-exist. We consider that each data source has a different importance and, hence, a priority. We formulate and solve a priority-aware model allocation optimization problem. Based on the structure of the optimal solution, we design a practical Priority-Aware Model-Distributed Inference (\ours)  algorithm that determines model allocation and distribution over devices by taking into account the priorities of different sources. Experiments were conducted on a real-life testbed of NVIDIA Jetson Xavier and Nano edge devices as well as in the Colosseum testbed with ResNet-50, ResNet-56, and GPT-2 models. The experimental results show that \ours performs priority-aware model allocation successfully while reducing the inference time as compared to baselines. 





\end{abstract}

%% file: intro.tex
\section{Introduction}
\label{sec:intro}

Data-distributed inference (DDI) is the traditional approach for Machine Learning (ML) models for fast inference. DDI distributes data to workers, which are comprised of edge servers, end users, and/or remote cloud (if available). An end user that would like to process input data offloads it to workers for inference. The end user itself could function as one of the workers by processing some of its own data.  Each worker keeps a pre-trained ML model, processes the offloaded data, and sends the output back to the end user. This approach, although very straightforward, has two disadvantages: (i) Communication cost is high, especially when the data size is large (e.g., high-resolution image); and (ii) Each worker should store the whole ML model, which puts a strain on end-user devices. 

\begin{figure}[t!]
    \centering
    \scalebox{0.195}{
    \includegraphics{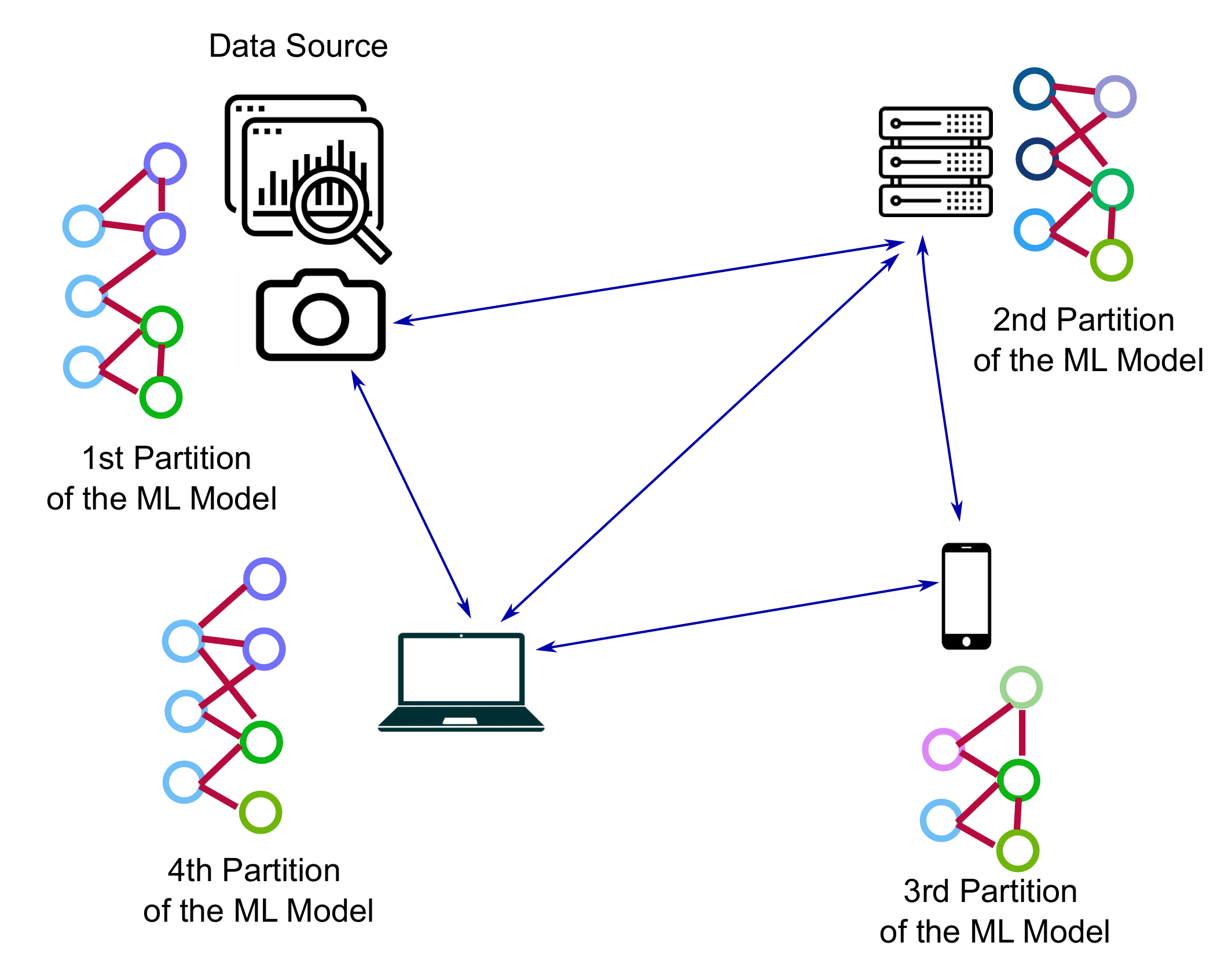}}
    \vspace{-10pt}
    \caption{Model-distributed inference and model parallelism.}
    \label{fig:MDI}
    \vspace{-20pt}
\end{figure}

\emph{Model-distributed} inference (also called \emph{model parallelism}) is emerging as a promising solution, where an ML model is partitioned and distributed across workers, Fig. \ref{fig:MDI}. The end user (data source), which has input data, may process a few layers of an ML model (1st partition in Fig. \ref{fig:MDI}) and transmit the feature vector of its last layer to a neighboring worker. The neighboring worker receives a feature vector and performs the calculations of the layers that are assigned to it (2nd partition in Fig. \ref{fig:MDI}). Finally, the worker that calculates the last layers of the ML model (4th partition in Fig. \ref{fig:MDI}) obtains and sends the output back to the end user (data source) that has the input data. We note that the workers perform parallel processing in this setup by pipelining. 
Thanks to parallelization, MDI reduces the inference time significantly. Furthermore, MDI is becoming more and more relevant in practice due to the sheer size of ML models (e.g., LLMs) and the difficulty of putting a whole model in one device.

The performance of MDI with heterogeneous resources is investigated in \cite{AR-MDI}, and an adaptive layer allocation mechanism across workers is designed. It is shown that MDI significantly reduces the inference time compared to data distributed inference when the data size is large \cite{AR-MDI}. MDI is applied to a multi-source scenario in \cite{li2023model}, where more than one worker in the system may have data to process using an ML model. 
%
%
The multi-source MDI \cite{li2023model} is efficient for fair resource allocation across multiple ML inference tasks, but it assumes that ML models and the datasets are the same at different sources and they have similar prioritization. 

\begin{figure}[t!]
    \centering
    \scalebox{0.21}{
    \includegraphics{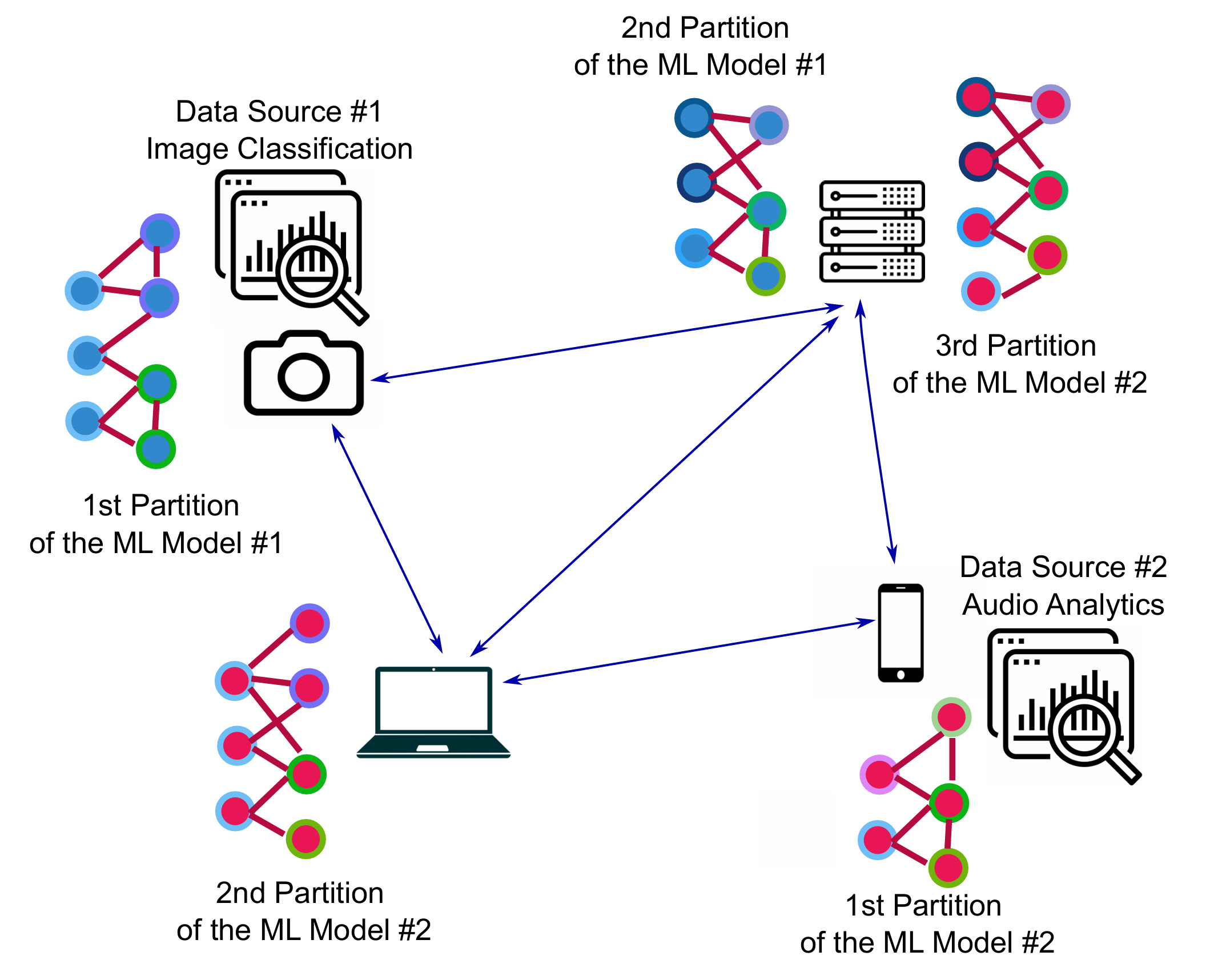}}
    \vspace{-5pt}
    \caption{Model-distributed inference for multiple data sources. 
    One of the data sources, either image classification or audio analytics, may have a higher priority in terms of processing their data.}
    \label{fig:PA-MDI}
    \vspace{-20pt}
\end{figure}

However, different sources could have different priorities depending on their applications in practical systems. For example, consider a camera surveillance system with multiple cameras at various locations in a house. One can expect that the video at the home's entry point may be more critical than others, so priority should be given to processing the video of this camera. Also, different sources may have different modalities, which is a plausible scenario in practice. For example, one device in the system may run an audio analytics ML model for better environment awareness, while another device may capture an image of a suspected situation and would like to analyze the image as soon as possible, as illustrated in Fig. \ref{fig:PA-MDI}. In this scenario, priority should be given to the image analysis as there is a suspected situation. We note that it is possible that multiple data sources and modalities may originate from the same device. 
As seen,  prioritization of one ML inference task to another may be needed when multiple data sources and modalities co-exist in an MDI setup. Thus, new techniques should be designed to embrace the differences and requirements of these various data sources. Our goal in this paper is to design a model-distributed inferencing mechanism that supports the prioritization of ML models.

We consider that each data source has a different importance and, hence, a priority in the system. Data sources with higher priorities should be processed faster than other data without hurting accuracy. Considering that ML models can be split into multiple partitions (tasks), our goal is to determine the best task allocation policy across the workers. In other words, we aim to address the following question: Which worker should process which partition of which data source at a given time? Our approach is to formulate an optimization problem to maximize ML inference accuracy while minimizing inference delay. Then, based on the structure of the optimization problem and its solution,  we design a practical Priority-Aware Model-Distributed Inference (\ours)  algorithm. 
We implement \ours in a real testbed consisting of NVIDIA Jetson Xaviers \cite{NVIDIA_Xavier} and Nanos 
\cite {NVIDIA_Nano} as well as in the Colosseum testbed \cite{colosseum} with ResNet-50, ResNet-56, and GPT-2 models. The experimental results show that \ours performs priority-aware model allocation successfully while reducing the inference time as compared to baselines. 






%% file: related.tex
\section{Related Work}
\label{sec:related}



Model distribution is initially considered for parallel deep neural network (DNN) training to overcome communication and storage limitations and provide scalability for DNNs. GPipe \cite{huang2019gpipe}  achieves close-to-linear speed up in DNN training. PipeDream \cite{harlap2018pipedream, harlap2019improving} eliminates idle workers and thus improves efficiency via a dedicated scheduling and batching mechanism. Weight prediction can be used to improve the performance of both GPipe and PipeDream \cite{chen2018efficient}. A resilient model-distributed training scheme that is robust to failing or severely straggling workers is designed in \cite{pengzhen2021respipe}. Model-distributed DNN training is designed for memory-constrained edge devices in \cite{li2021model}. As compared to this line of work, we consider model distribution for inference rather than training.

An earlier work on distributed inference is a two-part DNN partition \cite{teerapittayanon2017distributed}. Specifically, a DNN is split into two parts, where the first few layers are placed at the end user, while the remaining layers are located in the cloud. The processing in the end user and cloud could be iterative or parallel. In iterative processing, an end user processes a few layers, and if it is confident with the output, it terminates the processing. Otherwise, feature vectors are sent to the cloud for further processing at subsequent DNN layers. In parallel processing, layers are partitioned and assigned to the end user and edge server (or cloud), which operate in parallel \cite{kang2017neurosurgeon} and \cite{eshratifar2019jointdnn}. 

In order to take full advantage of edge networks, model distributed inference with multiple split points is developed. A model partitioning problem for MDI is formulated in \cite{EdgePipe-hu2021pipeline} using dynamic programming. An adaptive and resilient layer allocation mechanism is designed in \cite{AR-MDI} to deal with the heterogeneous and time-varying nature of resources. Model-distributed inference for multiple sources is developed in \cite{li2023model}. A Model-distributed inference mechanism is designed by considering early exit \cite{bolukbasi, branchynet, shallowdeep} in \cite{MDI-Exit}. As compared to this line of work, we focus on designing model-distributed inference for multiple modalities, where different modalities may have different priorities, which is not considered in the model-distributed inference literature.


We note that various other methods to reduce the memory, communication, and computation footprints of MLs have been proposed, including conditional architectures \cite{han2021dynamic},  pruning \cite{liu2020pruning,koyuncu:c32}, quantization \cite{chen2015compressing,gong2014compressing,han2015deep, gupta2015deep,courbariaux2014training}, and gradient sparsification \cite{wangni2018gradient, alistarh2018convergence}. Joint consideration of DNN bi-partitioning and pruning is considered in \cite{jankowski2020joint,shi2019improving}. 
Our work is complementary to these alternate techniques in the sense that the performance of \ours can be further improved using one or more of these methods.




%% file: model.tex
\section{System Model and Background}
\label{sec:system}
{\bf Setup.} We consider an edge network with end users and edge servers. We name these devices as workers. There are $N$ workers in the system, and the set of workers is $\Nset$. We name the $n$th worker as ``Worker $n$''. The set of all the nodes which are neighbors of Worker $n$ is denoted by $\Nset_n$. 
Some of these workers are also data sources, i.e., they may have one or more data modalities to process. 
%
The source workers would like to process their data as soon as possible by exploiting their own computing power as well as that of other workers. 

 
 

We consider a dynamic edge computing setup where workers join and leave the system anytime. Also, computing and communication delays of workers could be heterogeneous and vary over time.  Workers form an ad-hoc topology depending on their geographical locations, proximity to each other, and surrounding obstacles. The workers are connected to each other using WiFi in an ad-hoc mode. 


\textbf{Dataset, ML Models and Modalities.} We assume that there are $M$ data sources (either same or different modalities) in the network, where each data source has potentially different data and ML model for processing. Suppose that Worker $n$ is a data source for data source $m$. Worker $n$ may already have data or collect data from the environment in real-time. The $d$th vectorized data of source $m$ is $A_m^d \in \mathcal{R}^{u_m}$, where $u_m$ is the dimension of the input data of source $m$. We consider that there are $D_m$ data points that belong to source $m$.

We assume a pre-trained ML model $\Mset_m$ with $L_m$ layers is used for source $m$. The set of layers is determined by $\Lset_m = \{ l_m^1, \ldots, l_m^{L_m} \}$, where $l_m^l$ is the $l$th layer. The feature vector at the output of the $l$th layer is $a_m^l(d)$ for data $A_m^d$. The output of the ML model for source $m$ is $y_m \in \mathcal{R}^{v_m}$, where $v_m$ is the size of the output (i.e., the number of labels). 

We consider that each source has a different importance/priority in the system. Thus, we associate each source $m$ with a positive weight value $\gamma_m$, where larger $\gamma_m$ means higher priority should be given to the corresponding source in terms of processing and minimizing inference time. 



\textbf{Model Partitioning.} We consider that ML model $\Mset_m$ for source $m$ is partitioned to $K_m$ parts vertically (i.e., each partition is made between consecutive layers).\footnote{We note that our focus is only on vertical (layer-wise) partitioning in this paper for easy presentation. Extending our work to both vertical and horizontal partitioning (layer- and neuron-wise) is straightforward.} The layers of the $k$th partition for data point $A_m^d$ is defined as task $T_m^k(d)$. Considering that the first layer in task $T_m^k(d)$ is $\lambda_{m,b}^k \in \Lset_m$ and the last layer is  $\lambda_{m,e}^k \in \Lset_m$, task $T_m^k(d)$ is defined as processing all the layers in $\Lset_{m,k} = \{ \lambda_{m,b}^k, \lambda_{m,{b+1}}^k, \ldots, \lambda_{m,e}^k \}$ with input feature vector $a_{\lambda_{m,b}^k}(d)$. Since there are $K_m$ partitions in $\Mset_m$, the set of tasks is $\Tset_m(d) = \{T_m^1(d), \ldots, T_m^{K_m}(d)\}$ for data $A_m^d$.

\textbf{Model-Distribution.} All workers in the system are capable of processing the tasks that are assigned to them. In other words, if the task $T_m^k(d)$ is assigned to Worker $n$ for processing, it activates the layers in $\Lset_{m,k}$ of model $\Mset_m$ with input feature vector $a_{\lambda_{m,b}^k}(d)$. It calculates all the layers in  $\Lset_{m,k}$ and produces the output feature vector $a_{\lambda_{m,{e+1}}^k}(d)$. If $a_{\lambda_{m,e}^k + 1}(d)$ is the output of the last layer in $\Lset_{m,k}$, the process is terminated, and the output $a_{\lambda_{m,{e+1}}^k}(d)$ is transmitted back to the source worker. Otherwise, i.e., if $a_{\lambda_{m,{e+1}}^k}(d)$ is not the last output, then a new task $T_m^{k+1}(d)$ is generated with input feature vector $a_{\lambda_{m,{e+1}}^k}(d)$. This paper aims to determine which worker to process each task $T_m^{k+1}(d)$ while minimizing the inference time by considering the priorities of different sources.

%% file: M2MDI.tex
\section{\label{sec:PA-MDI} Priority-Aware Model-Distributed Inference}

In this section, we design a priority-aware model distributed inference framework. We first formulate an optimization problem to maximize ML inference accuracy while minimizing inference delay. Then, based on the structure of the optimization problem and its solution,  we design a practical Priority-Aware Model-Distributed Inference (\ours)  algorithm.

\subsection{Optimization Problem} Let us assume that $\pi_m^k(d)$ is the policy for task $T_m^k(d)$, where $\pi_m^k(d) \in \Nset$, which means that $\pi_m^k(d)$ is the worker that should process the task $T_m^k(d)$. Our goal is to maximize the accuracy of the processed data/tasks over all sources while minimizing the inference time by taking into account the prioritization of different sources.  

Let $I(\pi_m(d))$ be the accuracy contribution (improvement) after all tasks in $\Tset_m(d)$ are processed, where $\pi_m(d) = \{ \pi_m^1(d), \ldots,  \pi_m^{K_m}(d)\}$. We characterize $I(\pi_m(d))$ as
\begin{align}\label{eq:acc_m_d}
    I(\pi_m(d)) = \alpha_m(d) \prod_{k \leq K_m} (1 - P(\pi_m^k(d))),
\end{align} where $\alpha_m(d)$ is the amount of accuracy improvement when data $A_m^d$ is processed successfully, and $P(\pi_m^k(d))$ is the probability that the processing of task $T_m^k(d)$ is not successful, which may happen due to the dynamic setup; a worker may leave the system without completing the tasks that it is supposed to process or packet losses over wireless links. The accuracy term $I(\pi_m(d))$ is defined by weighting the maximum accuracy $\alpha_m(d)$ with task success probability $\prod_{k \leq K_m} (1 - P(\pi_m^k(d)))$. The task success probability has multiplication form as all the tasks in  $\Tset_m(d)$ should be processed for successful task inference for data $A_m^d$.




The total accuracy across all data sources and data points is
\begin{align} \label{eq:Ipi}
    I(\pi) = \sum_{m=1}^{M} \gamma_m (\sum_{d=1}^{D_m} I(\pi_m(d))),
\end{align} where $\pi = \{\pi_1, \ldots, \pi_M\}$ and $\pi_m = \{\pi_m(1), \ldots,  \pi_m(D_m) \}$. The $\gamma_m$ term in (\ref{eq:Ipi}) shows the importance of data source $m$. In particular, if a data source is important (higher $\gamma_m$), priority should be given to its tasks. Our goal is to maximize the cumulative accuracy $I(\pi)$ while minimizing the inference delay, which we formulate next. 

The cumulative inference delay $\Delta(\pi)$ under policy $\pi$ is characterized as 
\begin{align}
    \Delta(\pi) = \sum_{m=1}^{M} \sum_{d=1}^{D_m} \sum_{k=1}^{K_m} \rho(\pi_m^k(d)), 
\end{align} where $\rho(\pi_m^k(d))$ is the inference delay of task $T_m^k(d)$ under policy $\pi_m^k(d)$. The inference delay consists of two parts: (i) computation delay for processing task $T_m^k(d)$ at Worker $\pi_m^k(d)$; and (ii) communication delay if task  $T_m^k(d)$ is offloaded to Worker  $\pi_m^k(d)$ noting that $\pi_m^k(d) \in \Nset$. The communication delay captures multi-hop transmission delay if task offloading is through multiple hops. 


Our goal is to maximize total accuracy while minimizing the inference delay, 
which is equivalent to 
\begin{align} \label{eq:opt}
    \max_{\pi, \beta} J(\pi) = \max_{\pi, \beta} \{ I(\pi) - \beta \Delta(\pi)  \}, 
\end{align} where $\beta$ is a Lagrange multiplier.

\subsection{Solution of the Optimization Problem}
Noting that (\ref{eq:opt}) is a convex optimization problem, and the optimization problem can be decomposed, i.e., $J(\pi)$ can be expressed as 
\begin{align}
    J(\pi) = & \sum_{m=1}^{M} \sum_{d=1}^{D_m} (  \gamma_m \alpha_m(d) \prod_{k \leq K_m} (1 - P(\pi_m^k(d))) - \nonumber \\
    & \beta \sum_{k=1}^{K_m} \rho(\pi_m^k(d)) ).  
\end{align} Minimizing $J(\pi)$ corresponds to determining the set of policies $\pi$ by choosing the minimum Lagrange multiplier $\beta$. Hence, minimizing $J(\pi)$ 
reduces to solving 
\begin{align} \label{eq:lagrange}
     \min_{\pi_m(d)} \frac{\sum_{k=1}^{K_m} \rho(\pi_m^k(d))}{\gamma_m \alpha_m(d) \prod_{k \leq K_m} (1 - P(\pi_m^k(d)))}  
\end{align} for each source $m$ and data $d$. In practice, a new task $T_m^k(d)$ is generated if $k=1$, i.e., if it is the first partition, or the previous task $T_m^{k-1}(d)$ is successfully processed. In this case, (\ref{eq:lagrange}) reduces to 
\begin{align} \label{eq:lagrange-red}
     \min_{\pi_m^k(d)} \frac{ \rho(\pi_m^k(d))}{\gamma_m \alpha_m(d)}. 
\end{align} In the next section, we describe how we implement this solution in a decentralized practical edge computing system.

\subsection{Design of \ours} 

Worker $n$ constructs a queue of tasks $\Hset_n$, which includes all the tasks that Worker $n$ is supposed to handle either by processing them or offloading them to one of its neighbors. Worker $n$ runs two parallel threads; (i) task processing (summarized in Alg. \ref{alg:PAMDI-process}), and (ii) communication and message passing (summarized in Alg. \ref{alg:PAMDI-msg}). Next, we explain the details of these threads.


\begin{algorithm} [t] 
	\caption{\ours task processing at Worker $n$} 
	\label{alg:PAMDI-process} 
	\begin{algorithmic}[1]            
            \STATE Initialize neighbors list $\tilde{\mathcal{N}}_n$ = $\mathcal{N}_n$. Update $F_n$ and $Q_n$. Send status requests to all neighbors $j \in \mathcal{N}_n$. 
            \STATE Receive $F_j$ and $Q_j$ from  neighbors $j \in \mathcal{N}_n$. Evaluate $d_{n,j}$.
            \STATE Fetch the highest priority and oldest task $T_m^k(d)$ from $\Hset_n$ according to $\gamma_m$ and $\delta(T_m^k(d))$.
            \WHILE{ $ \tilde{\mathcal{N}}_n \neq \varnothing $ }
            \STATE Solve  $ j^* = \arg \min_{j \in \Nset_n} \{ \frac{d_{n,j} + \delta(T_m^k(d)) + F(T_m^k(d)) F_j + Q_j}{\gamma_m \alpha_m(d)}\}$.
            \IF{$j^* == n$}
                \STATE Process $T_m^k(d)$ locally and obtain  $a_{\lambda_{m,e}^{k+1}}(d)$.
                \IF{$T_m^k(d)$ is the last task of $A_m^d$ AND Worker $n$ is the source of $m$}                    
                    \STATE Output the result.
                    \STATE Create a new task $T_m^1(d+1)$ for data point $A_m^{d+1}$, and insert it into $\Hset_n$.
                \ELSIF {$T_m^k(d)$ is the last task of $A_m^d$ AND Worker $n$ is NOT the source of $m$}
                    \STATE Send the output vector to the source of $m$.       
                \ELSE
                \STATE Create $T_m^{k+1}(d)$ and insert in $\Hset_n$. 
                \ENDIF
            \ELSE
                \STATE Send RTC message to $j^*$.  
                \IF{CTC is received} \STATE Send $a_{\lambda_{m,b}^k}(d)$ to $j^*$.
                \ELSE
                    \STATE $\tilde{\mathcal{N}}_n = \tilde{\mathcal{N}}_n - j$.
                \ENDIF
            \ENDIF
            \ENDWHILE 
	\end{algorithmic} 
\end{algorithm}


\textbf{Task processing and offloading.} In \ours, while there exists a task in $\Hset_n$ at Worker $n$, it runs Alg. \ref{alg:PAMDI-process}. All workers measure their computing performance through floating-point operations per second (FLOPS), which is $F_n$ for Worker $n$. They also estimate how long it takes to finish the tasks that are already assigned to them, which is $Q_n$ for Worker $n$. Worker $n$ sends its $F_n$ and $Q_n$ data to their neighbors upon their neighbors' requests. Each task $T_m^k(d)$ in $\Hset_n$ has age information $\delta(T_m^k(d))$, which shows the lifetime of task $T_m^k(d)$ after its creation. The age information captures communication delay if this task has already been offloaded as well as waiting time in task queues. 


Worker $n$ fetches the highest priority task from its queue $\Hset_n$. The oldest task considering $\delta(T_m^k(d))$ is selected among the equal priority tasks. Let us assume this task is  $T_m^k(d)$.  For this task, it determines 
\begin{align} \label{eq:prior}
    j^* = \arg \min_{j \in \Nset_n} \{ \frac{d_{n,j} + \delta(T_m^k(d)) + F(T_m^k(d)) F_j + Q_j}{\gamma_m \alpha_m(d)}  \}, 
\end{align} where $d_{n,j}$ is the communication delay between Workers $n$ and $j$ (noting $d_{n,n}=0$), $F(T_m^k(d))$ is the number of floating-point operations needed by task $T_m^k(d)$, and $F_j$ is the computing performance that node $j$ can offer to task $T_m^k(d)$. We note that the numerator of (\ref{eq:prior}), \ie $d_{n,j} + \delta(T_m^k(d)) + F(T_m^k(d)) F_j + Q_j$ corresponds to $\rho(\pi_m^k(d))$ in (\ref{eq:lagrange-red}).

Worker $n$ checks if $j^* = n$. If so, Worker $n$ processes task $T_m^k(d)$ locally and creates a new task $T_m^{k+1}(d)$ or terminates the task if it is the last task for data $A_m^d$. If $j^* \neq n$, then Worker $n$ transmits the feature vector $a_{\lambda_{m,b}^k}(d)$ associated with task $T_m^k(d)$ to Worker $j^*$. Worker $j^*$ solves (\ref{eq:prior}) for its neighbors and decides whether to offload task $T_m^k(d)$ to its neighbors or process locally. 

If Worker $n$ is a source node of  $m$, and if it just finished a task for data point $A_m^d$ either by offloading or processing all its tasks associated with $A_m^d$, then it creates a new task  $T_m^1(d+1)$ for data point $A_m^{d+1}$.


We should note that multiple workers may solve (\ref{eq:prior}) to offload their respective tasks. If more than one worker selects $j^*$ around the same time, worker $j^*$ is overloaded. To avoid this, we implement a request to compute (RTC) and clear to compute (CTC) mechanism inspired by RTS/CTS mechanism of CSMA/CA \cite{567423}. In our RTC/CTC mechanism, Worker $n$ sends an RTC message to Worker $j^*$ about its intention of offloading its task along with how much computing (floating point operations) the task requires. Worker $j^*$ sends a CTC message to all its neighboring workers. The CTC message includes which worker can transmit (assume Worker $n$) and how much computing it will need. Then, the selected Worker $n$ offloads its task to Worker $j^*$.

\begin{algorithm} [t!] 
	\caption{Communication and message passing operation of \ours at Worker $n$} 
	\label{alg:PAMDI-msg} 
	\begin{algorithmic}[1]
        \STATE Listen for the messages from neighbors $\Nset_n$. 
        \IF {$msg_j$ is received from neighbor $j \in \Nset$}
            \IF{$msg_j$ is $a_{\lambda_{m,b}^k}(d)$}
                \STATE Insert $T_m^k(d)$ into $\Hset_n$.
            \ELSIF{$msg_j$ is status request}
                \STATE Transmit $F_n$ and $Q_n$ to Worker $j$.
            \ELSIF{$msg_j$ is RTC AND Worker $n$ is not processing a task}
                    \STATE Transmit CTC to all $j \in \Nset_n$.
            \ELSIF{$msg_j$ is output vector}
                \STATE Create a new task $T_m^1(d+1)$ for data point $A_m^{d+1}$, and insert it into $\Hset_n$.
            \ENDIF
            \ENDIF
	\end{algorithmic} 
\end{algorithm}

\textbf{Communication and message passing.} In parallel to the task processing and offloading thread (Alg. \ref{alg:PAMDI-process}), Worker $n$ runs the communication and message passing thread, which is summarized in Alg. \ref{alg:PAMDI-msg}. When a message $msg_j$ is received from neighboring node $j$, Worker $n$ checks the kind of the message. If the received message is $a_{\lambda_{m,b}^k}(d)$, it means that a task is offloaded to Worker $n$, so it creates a new task $T_m^k(d)$ and inserts it into $\Hset_n$. If it is a status request, Worker $n$ transmits $F_n$ and $Q_n$ to Worker $j$. If is an RTC message and Worker $n$ does not process a task, it sends a CTC message to all of its neighbors. Finally, if the message is an output vector regarding the computation of Worker $n$'s data point $A_m^d$, it creates a new task $T_m^1(d+1)$ for data point $A_m^{d+1}$, and insert it into $\Hset_n$.

\begin{figure}[t]
\centering
\centerline{\includegraphics[clip, trim=0cm 0cm 0cm 0cm, width=0.5\textwidth]{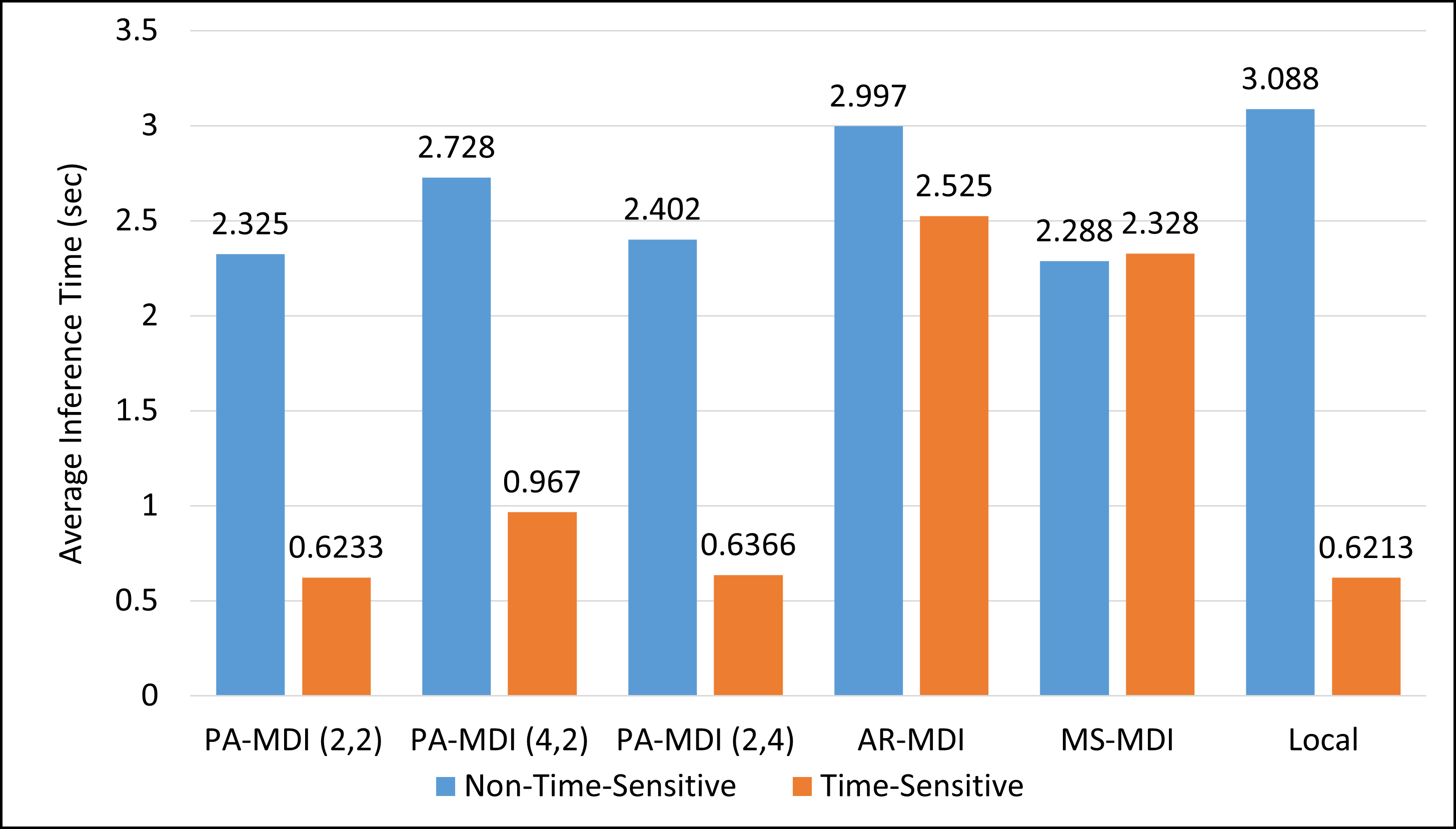}}
\caption{Worker $A$, who hosts ``Non-Time-Sensitive'' data, has dataset CIFAR-10 (224x224) which is processed by ResNet-50, while Worker $D$, who hosts ``Time-Sensitive'' data, has dataset CIFAR-10 (32x32) which is processed by ResNet-56. 
}
\label{fig:test1}
\end{figure}


%% file: results.tex

\begin{figure}[htbp]
\centering
\centerline{\includegraphics[clip, trim=0cm 0cm 0cm 0cm, width=0.5\textwidth]{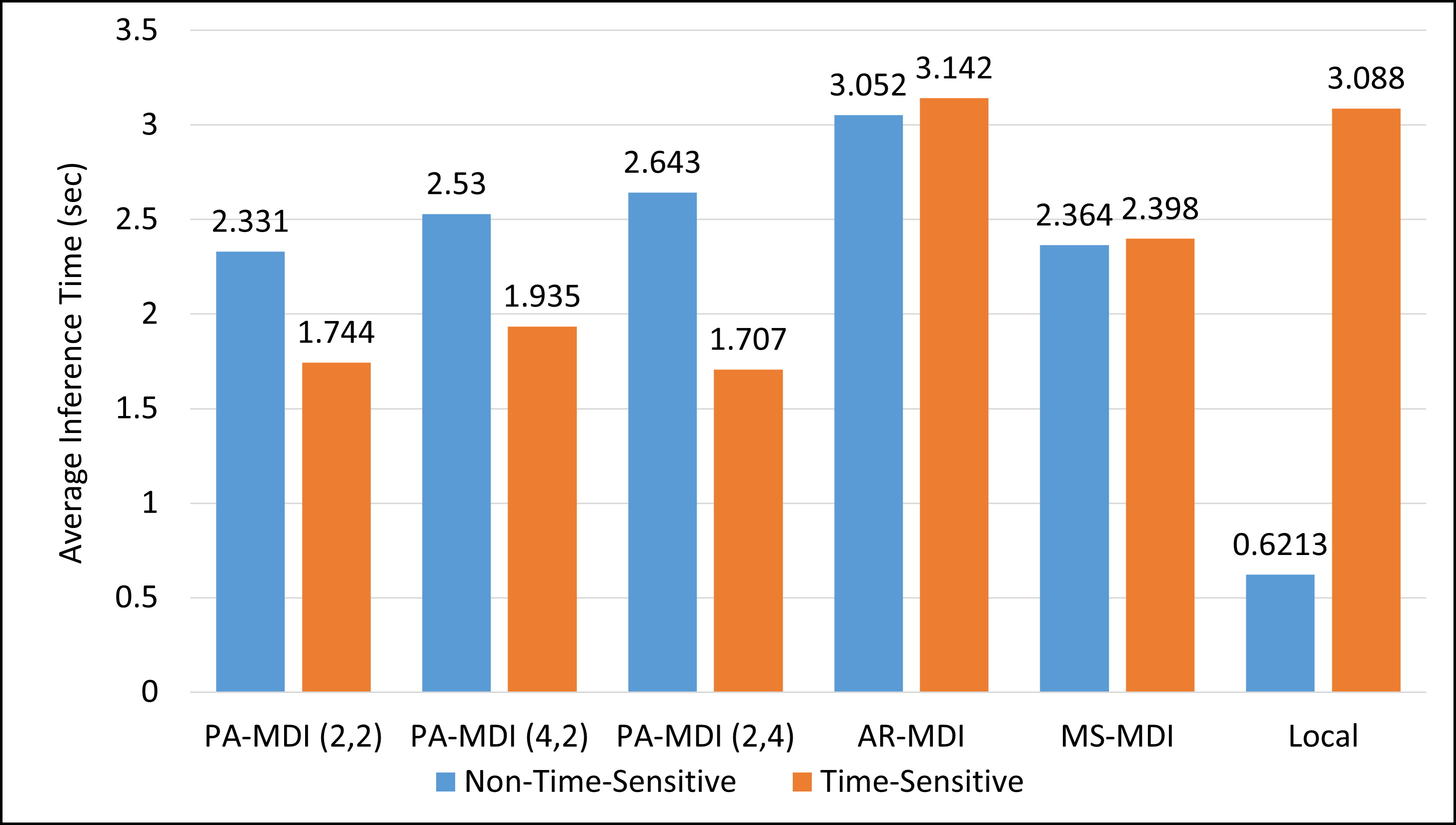}}
\caption{Worker $A$ has a small data set (CIFAR-10 (32x32), which is processed by distributed ResNet-56) while Worker $D$ has a larger dataset (CIFAR-10 (224x224), which is processed by distributed ResNet-50). Worker $A$ hosts ``Non-Time-Sensitive'' data, while Worker $D$ hosts ``Time-Sensitive'' data.  
}
\label{fig:test2}
\end{figure}

\section{Experimental Results}
\label{sec:results}
In this section, we evaluate the performance of our algorithm \ours in a real testbed consisting of NVIDIA Jetson Xaviers \cite{NVIDIA_Xavier} and Nanos \cite{NVIDIA_Nano} as well as in the Colosseum testbed \cite{colosseum} with ResNet-50, ResNet-56, and GPT-2 models. For each set of experiments, we will first describe the testbed architecture, DNN models, datasets, and baselines. We will then provide the corresponding experimental results.


\subsection{\label{sec:resFullConNVIDIA}Image Classification on NVIDIA Jetsons}

\subsubsection{Setup}

\textbf{Testbed.} To conduct our experiments, we utilized the NVIDIA Jetson Xavier \cite{NVIDIA_Xavier}, a commonly used platform for ML development on edge devices. The Jetsons are equipped with a 6-core NVIDIA Carmel CPU @ 1900 MHz and 16 GB of RAM.  We use five Jetsons in our experiments (Workers $A$, $B$, $C$, $D$, and $E$), and all these devices are connected to each other, forming a mesh topology. The Jetsons are connected using WiFi in an ad-hoc mode, i.e., there are no access points. The shared bandwidth is measured at around 20Mbps. We use the PyTorch deep learning framework (CPU-version), where computation and communication tasks are parallelized.

\textbf{Datasets and DNN Models.} We use CIFAR-10 dataset \cite{cifar_dataset} utilizing 10,000 test images.  We use ResNet-50 \cite{Resnet50} and ResNet-56 \cite{Resnet56} as ML inference models. The input to ResNet-50 model is CIFAR-10 data resized to 224x224, while the input to ResNet-56 is 32x32 CIFAR-10 data. We consider two types of data sources in our experiments; ``Time-Sensitive'' and ``Non-Time-Sensitive''. We set the prioritization parameter $\gamma_m$ of the ``Time-Sensitive'' data source to a large positive number. We consider the accuracy parameters $\alpha_m(d)$ equal to each other for different data sources and data points. We consider three different scenarios where smaller or larger CIFAR-10 data could be ``Time-Sensitive''. Worker $A$ is the source for all ``Non-Time Sensitive'' tasks, while Worker $D$ is the source for all ``Time-Sensitive'' tasks. 

DNN models are partitioned roughly uniformly for \ours. For example, ResNet50 has 23 blocks. If we split it into two parts (tasks), the first task will contain 12 blocks, while the second task will have 11 blocks. We evaluate \ours for different partitions; ``\ours $(\mu,\eta)$'' corresponds to the scenario that the ML model of Time-Sensitive data source is split into $\mu$ partitions and the ML model of Non-Time-Sensitive data source is split into $\eta$ partitions.
 

\textbf{Baselines.} We evaluate \ours as compared to the following baselines. 
\begin{itemize}
    \item ``AR-MDI'' \cite{AR-MDI}.  This is an adaptive and resilient MDI algorithm designed for a single-source and circular topology.  DNN models are distributed over five workers, where Workers $A$, $B$, $C$, $D$, and $E$ form a circular topology. Two sources are Workers $A$ and $D$. 
    \item ``MS-MDI'' \cite{li2023model}, which is an extension of AR-MDI to handle multiple sources, but with same priorities. It considers a circular topology of Workers $A$, $B$, $C$, $D$, and $E$, where  Workers $A$ and $D$ are sources. 
    \item ``Local'', where each source node processes its tasks locally. Thus, there is no model distribution or task offloading in this baseline. 
    %
%
\end{itemize}

\textbf{Performance Metric.} We use ``Average Inference Time'' as a performance metric. If Worker $n$ is a source for data $m$, it records the task creation and completion times for each of its tasks $T_m^k(d)$. The difference between task completion and creation times is $\tau_m^k(d)$. We sum $\tau_m^k(d)$ across all tasks and obtain $\tau_m(d) = \sum_{k=1}^{K_m}\tau_m^k(d)$. We take the average of $\tau_m(d)$ across all data points to report the average inference time for source $m$.

\begin{figure}[htbp]
\centering
\centerline{\includegraphics[clip, trim=0cm 0cm 0cm 0cm, width=0.5\textwidth]{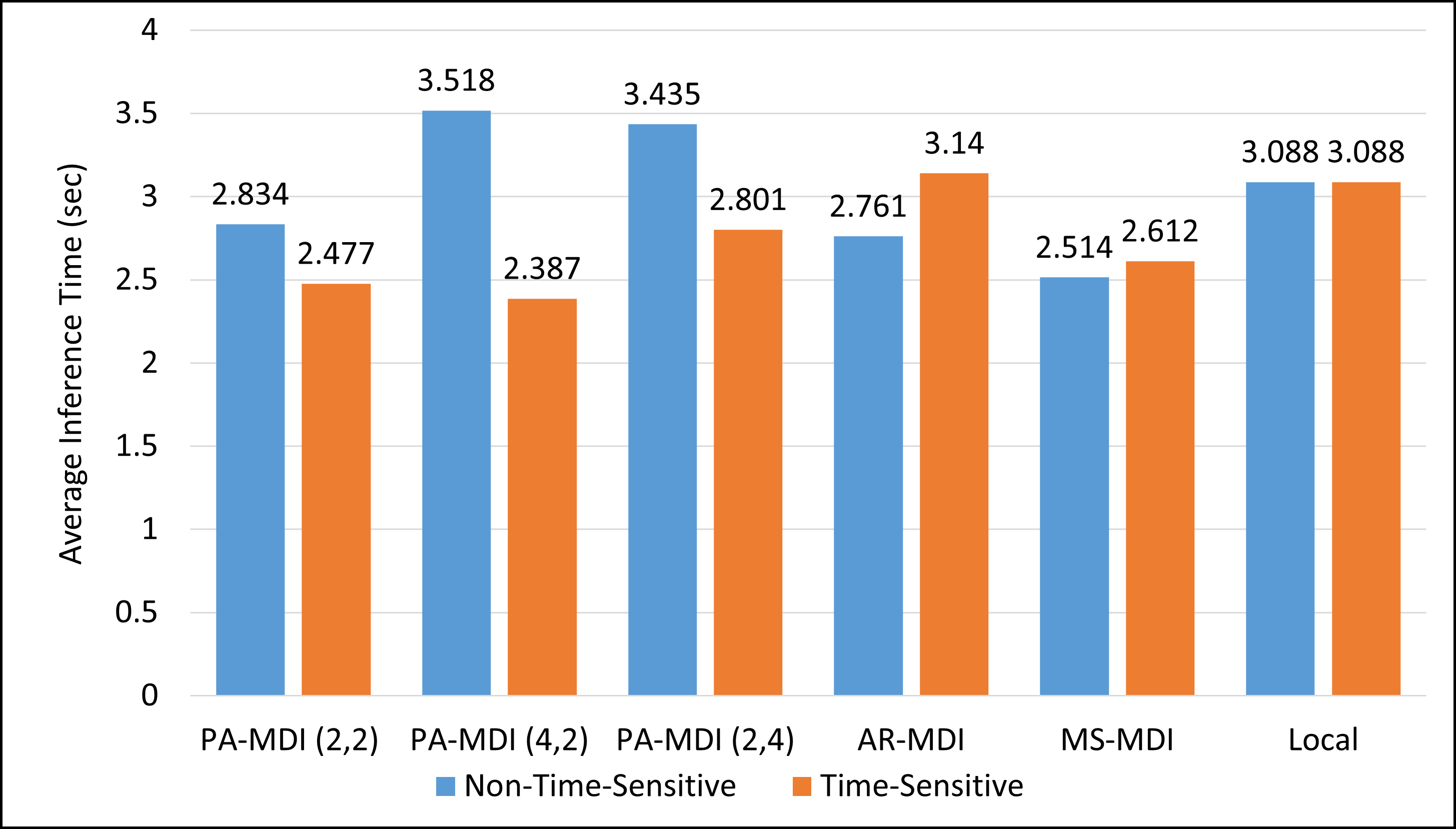}}
\caption{Both Workers $A$ and $D$ have larger dataset (CIFAR-10 (224x224), which is processed by ResNet-50). Similar to the previous scenarios, Worker $A$ hosts ``Non-Time-Sensitive'' data, while Worker $D$ hosts ``Time-Sensitive'' data.
}
\label{fig:test3}
\vspace{-10pt}
\end{figure}

\subsubsection{Results} We evaluate the performance of \ours for three different scenarios. In the first scenario, Worker $A$, who hosts ``Non-Time-Sensitive'' data, has a larger dataset; CIFAR-10 (224x224), which is processed by ResNet-50, while Worker $D$, who hosts ``Time-Sensitive'' data, has a small dataset; CIFAR-10 (32x32), which is processed by ResNet-56. The average inference time for \ours (2,2), \ours (4,2), \ours (2,4) as compared to AR-MDI, MS-MDI, and Local is provided in Fig. \ref{fig:test1}. 
%
%
\ours improves the average inference time compared to both AR-MDI and MS-MDI because AR-MDI does not consider multiple sources, and MS-MDI does not consider different data sources and their corresponding prioritization.  
In other words, AR-MDI does not consider multi-source optimization, which leads to unnecessary congestion over communication links and inefficient usage of computational resources. MS-MDI improves over AR-MDI but lacks mechanisms to prioritize ``Time-Sensitive'' data sources. Compared to AR-MDI and MS-MDI, our approach reduces the average inference time of the ``Time-Sensitive'' data up to 75.3\% and 73.2\%, respectively. 

The average inference times of ``Time-Sensitive'' data source are similar in both \ours and Local algorithms in Fig. \ref{fig:test1} because the ``Time-Sensitive'' data source is the small dataset in this scenario. Thus, the best approach is to process tasks locally. As seen, \ours algorithm approaches the best solution in this scenario. Also, \ours improves over Local in ``Non-Time-Sensitive'' data source by 24.7\%, thanks to model-distribution and parallelization. 

Fig. \ref{fig:test1} also shows that if the number of partitions (tasks) of ``Non-Time-Sensitive'' data source is larger than that of ``Time-Sensitive'' data source (for example, \ours (4,2) scenario), this negatively affects the performance of \ours. The reason is that more non-time-sensitive tasks create congestion in the network, so it becomes more challenging to prioritize time-sensitive data sources. This shows the importance of arranging the number of partitions depending on the importance of data sources.




In the second scenario, Worker $A$ has a small data set (CIFAR-10 (32x32), which is processed by distributed ResNet-56) while Worker $D$ has a larger dataset (CIFAR-10 (224x224), which is processed by distributed ResNet-50). Worker $A$ hosts ``Non-Time-Sensitive'' data source, while Worker $D$ hosts ``Time-Sensitive'' data source. As shown in Fig. \ref{fig:test2}, \ours reduces the average inference time of time-sensitive tasks over AR-MDI and MS-MDI by 45.7\% and 28.8\%, respectively, which is significant. \ours also shows significant improvement over Local for time-sensitive tasks thanks to its model distribution and prioritization. 


In the third scenario, both Workers $A$ and $D$ have a larger dataset (CIFAR-10 (224x224), which is processed by distributed ResNet-50). Similar to the previous scenarios, Worker $A$ hosts ``Non-Time-Sensitive'' data source, while Worker $D$ hosts ``Time-Sensitive'' data source.
Fig. \ref{fig:test3} shows that our approach reduces the average inference time of the ``Time-Sensitive'' source by up to 24.0\%, 8.6\%, and 22.7\% as compared to AR-MDI, MS-MDI, and Local, respectively, thanks to data prioritization and model distribution.


\begin{figure}[htbp]
\centering
\centerline{\includegraphics[clip, trim=0cm 0cm 0cm 0cm, width=0.48\textwidth]{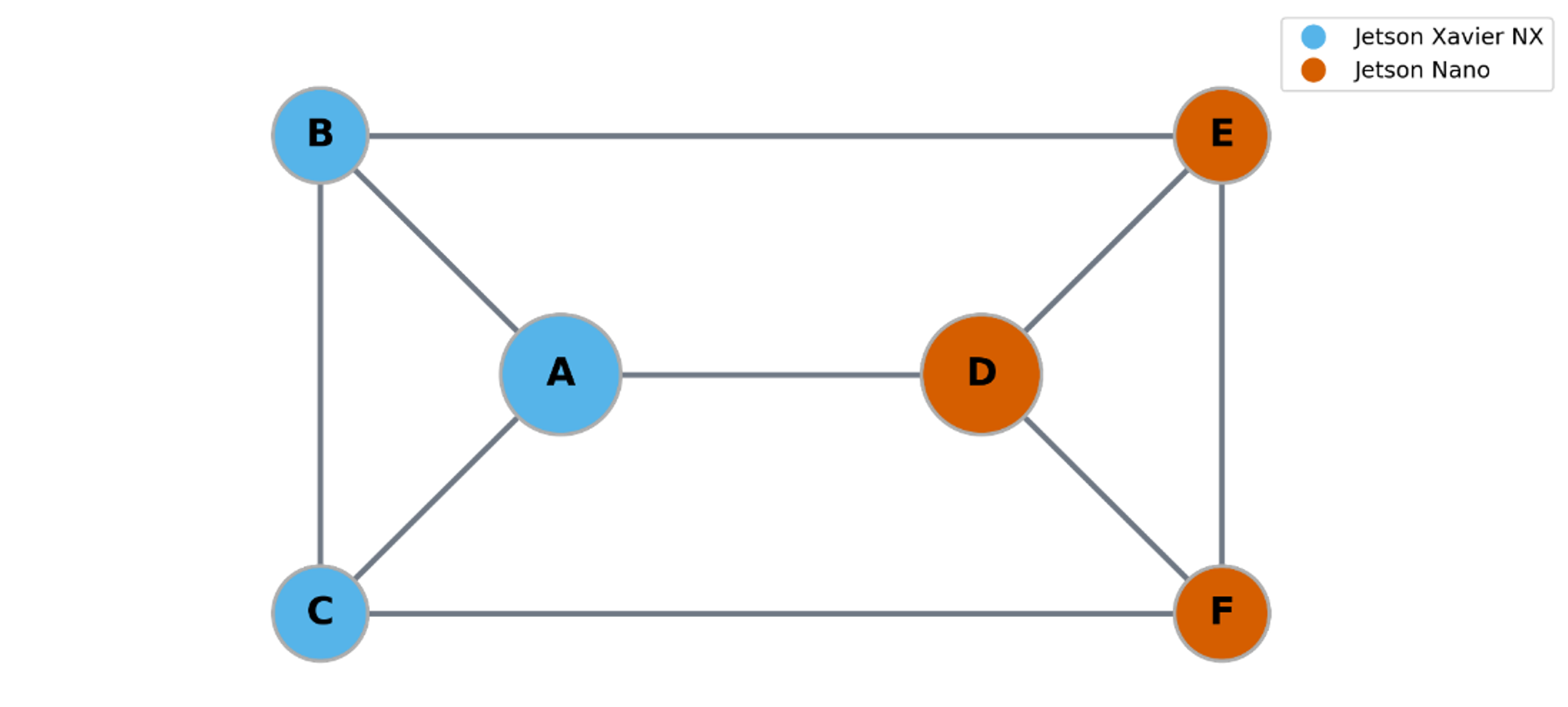}}
\vspace{-10pt}
\caption{Multi-hop topology.}
\label{fig:multi-topo}
\vspace{-15pt}
\end{figure}

\subsection{Heterogeneous and Multi-Hop Setup with NVIDIA Jetsons}

\subsubsection{Setup}  We created a multi-hop topology of heterogeneous NVIDIA Jetsons. The topology shown in Fig. \ref{fig:multi-topo} consists of three Jetson Xavier devices and three Jetson Nano devices. These devices have different computing capabilities, where Jetson Nano has a less powerful CPU and smaller RAM, equipped with a 4-core ARM Cortex A57 CPU @ 1430 MHz and 4 GB of RAM. All Jetsons are connected via Wi-Fi in an ad-hoc mode, similar to the fully connected topology in Section \ref{sec:resFullConNVIDIA}.

The experiment uses the ResNet-50 model and a modified CIFAR-10 dataset (re-sized to 224x224). Workers $A$ and $D$ serve as source nodes in Fig. \ref{fig:multi-topo}. 
We consider that the baselines AR-MDI and MS-MDI, as they are designed to operate over circular topologies, consider the following chains of nodes in this topology: (i) AR-MDI uses $[ A, B, E, D, F, C ]$ and $[ D, F, C, A, B, E ]$ and (ii) MS-MDI uses $[ A, B, E, D, F, C ]$ and $[ D, F, C, A, B, E ]$  for sources $A$ and $D$, respectively.  


\subsubsection{Results}

\begin{figure}[htbp]
\centering
\centerline{\includegraphics[clip, trim=0cm 0cm 0cm 0cm, width=0.48\textwidth]{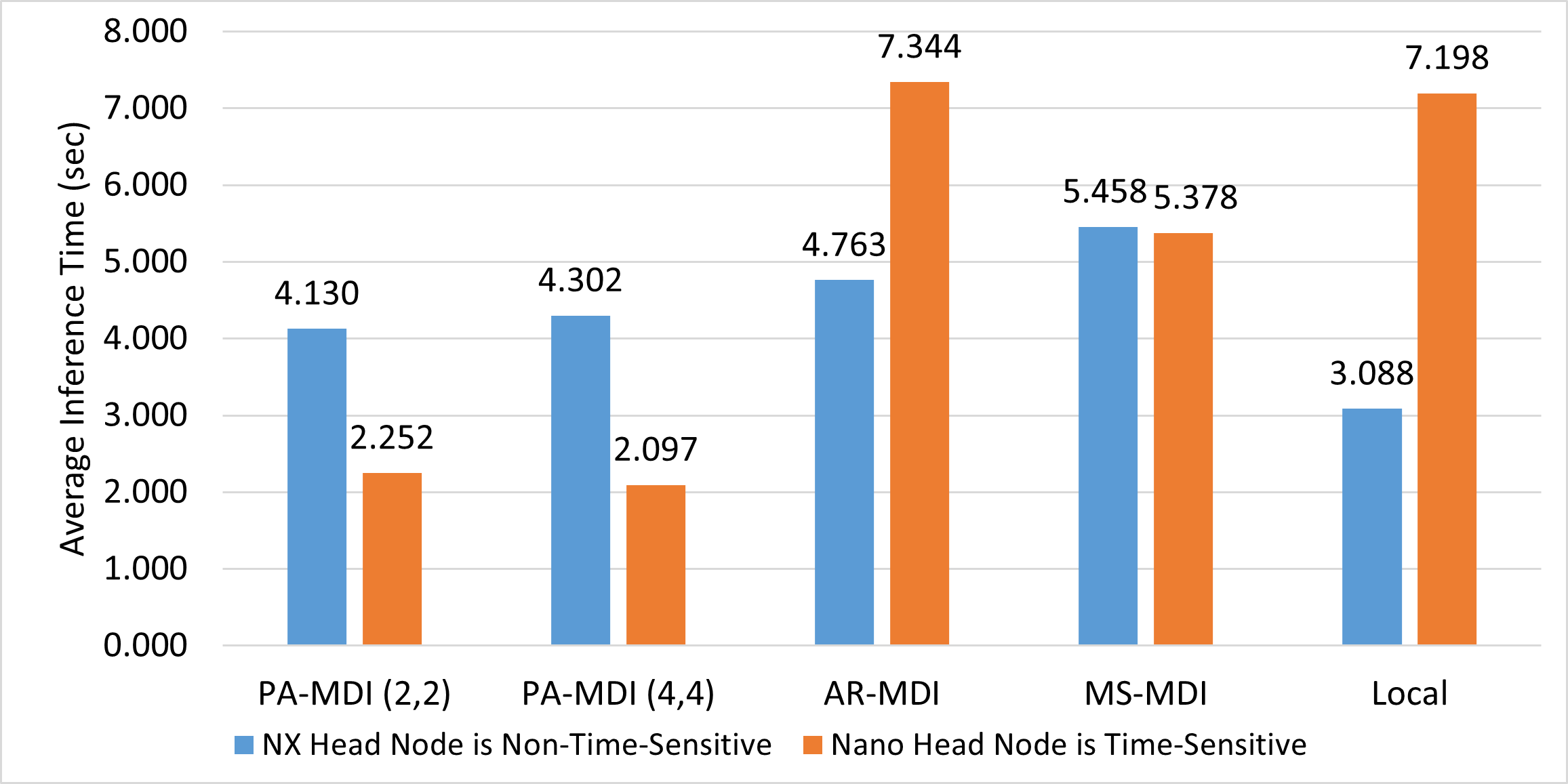}}
\caption{Average inference time for the multi-hop topology in Fig. \ref{fig:multi-topo}. Worker $A$, a Jetson Xavier device, hosts the ``Non-Time-Sensitive'' data source, while Worker $D$, a Jetson Nano device, hosts the ``Time-Sensitive'' data source.
}
\label{fig:test-6nodes-static-A0-D1}
\vspace{0pt}
\end{figure}

In the first scenario, Worker $A$, a Jetson Xavier source node, hosts the ``Non-Time-Sensitive'' data source, while Worker $D$, a Jetson Nano, hosts the ``Time-Sensitive'' data source. Fig. \ref{fig:test-6nodes-static-A0-D1} shows that \ours significantly reduces the average inference time of ``Time-Sensitive'' tasks compared to the AR-MDI, MS-MDI, and Local baselines thanks to considering different priorities of different sources.
In particular, compared to AR-MDI, MS-MDI, and Local, \ours reduces the average inference time for ``Time-Sensitive'' data source by up to 71.4\%, 61.0\% and 70.1\%, respectively. \ours also reduces the inference time for ``Non-Time-Sensitive'' data compared to AR-MDI and MS-MDI thanks to its ability to operate over multi-hop topology and efficient resource allocation.

\begin{figure}[htbp]
\centering
\centerline{\includegraphics[clip, trim=0cm 0cm 0cm 0cm, width=0.48\textwidth]{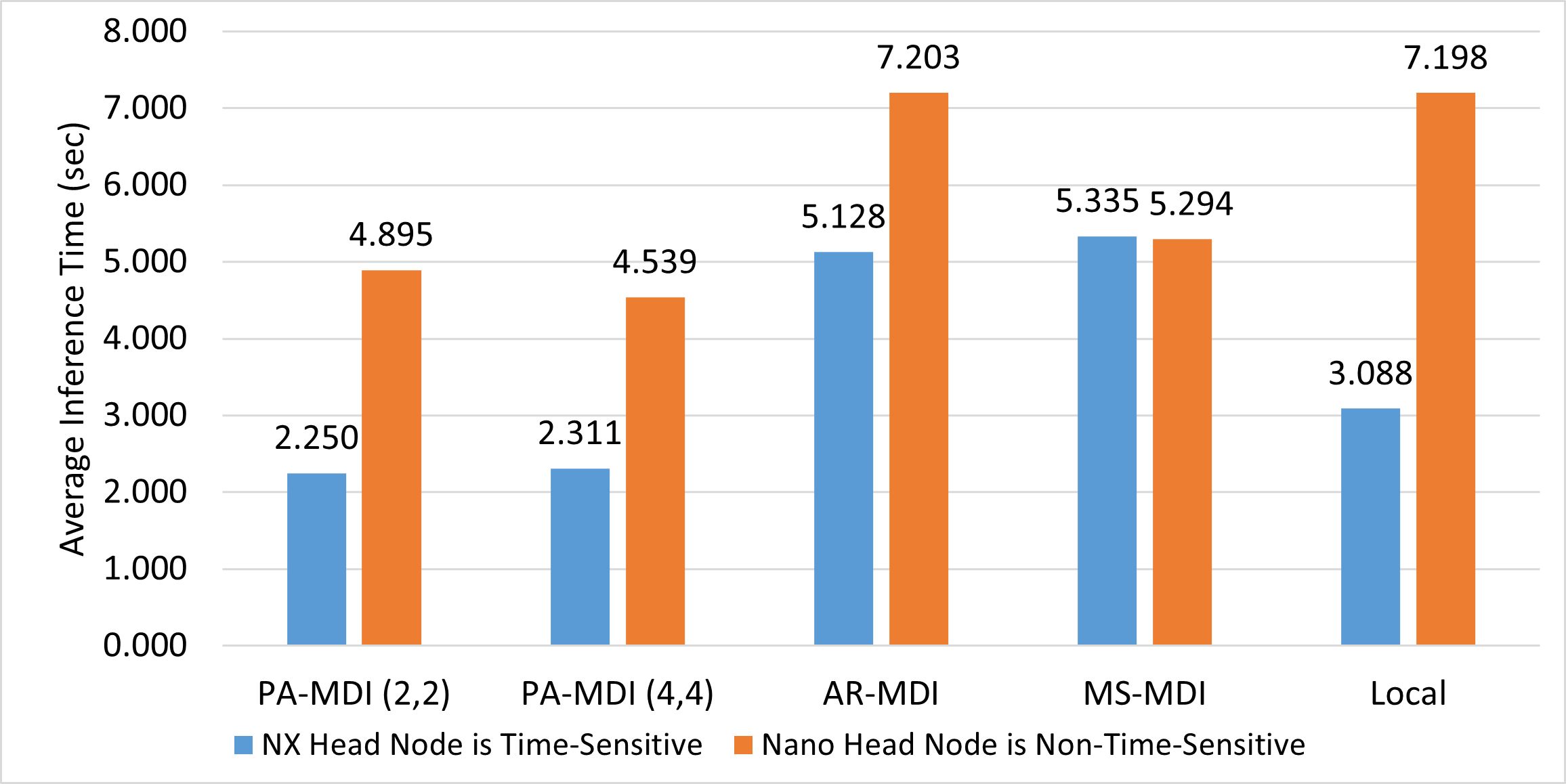}}
\caption{Average inference time for the multi-hop topology in Fig. \ref{fig:multi-topo}. Worker $A$, a Jetson Xavier device, hosts the ``Non-Time-Sensitive'' data source, while Worker $D$, a Jetson Nano device, hosts the ``Time-Sensitive'' data source.
}
\label{fig:test-6nodes-static-A1-D0}
\vspace{0pt}
\end{figure}

In the second scenario, Worker $A$, a Jetson Xavier device, hosts the ``Time-Sensitive'' data source, while Worker $D$, a Jetson Nano device, hosts the ``Non-Time-Sensitive'' modality. Fig. \ref{fig:test-6nodes-static-A1-D0} shows similar results to those observed in the previous scenario. \ours significantly reduces the average inference time for the ``Time-Sensitive'' tasks as compared to the AR-MDI, MS-MDI, and Local baselines. \ours improve the average inference time of ``Time-Sensitive'' data sources by up to 56.1\%, 57.8\%, and 27.1\% as compared to AR-MDI, MS-MDI, and Local, respectively.  \ours also improves the average inference time of ``Non-Time-Sensitive'' sources thanks to its very design of operating and resource allocation over multi-hop topologies. 


\subsection{GPT-2 on Colosseum Testbed}

We evaluate the performance of \ours algorithm on a larger testbed; Colosseum \cite{colosseum} with more capable devices as compared to NVIDIA Jetsons. We focus on evaluating the performance of \ours for large language models (LLMs), which have demonstrated remarkable capabilities in natural language processing. In particular, we evaluate LLMs in \ours, where the pre-trained GPT-2 model \cite{radford2019language} is used. 

\subsubsection{Setup}

\textbf{Colosseum Platform.}  Colosseum is an open and public large-scale wireless testbed \cite{bonati2021colosseum}. It provides multiple powerful and computationally capable nodes to meet the demands of complex model computations. 
We used five Standard Radio Nodes (SRNs) of Colosseum as computation nodes. Each SRN is a state-of-the-art server equipped with 46 available cores from Intel Xeon E5-2650 CPUs and an accessible NVIDIA Tesla K40m GPU, along with 120GB of memory for user applications. It is also noted that ML inference in our project is performed solely by CPUs.

The network scenarios we used are intra-network named Collaboration Network in Colosseum, which operates on a virtual network based on 10Gb Base-T devices. The topology remains the same as in the previous fully-connected Jetson-based testbed in section \ref{sec:resFullConNVIDIA}. Thus, all five nodes are interconnected using a mesh topology. 

\textbf{GPT-2 and Inputs.} 
All nodes host GPT-2 model. GPT-2 is a large language model based on transformers developed by OpenAI. Hugging Face provides different configurations and scales of the pre-trained GPT-2 model \cite{GPT2URL}. We selected the smaller English GPT-2 model from them, which has 12 layers, 768 hidden units, 12 heads, and 117 million parameters. The model inference is backed by CPU-version PyTorch and Hugging Face Transformers library in our experiments. We consider that models are partitioned roughly uniformly by the number of layers. For instance, if the model is divided into two parts (tasks), the first part would have 6 layers, while the second part would contain the rest 6 layers.

The model inputs are randomly generated tokens. Each batch of tokens is considered as a task, with a fixed sequence length of 64 and a hidden size of 768. At the same time, the batch size is variable, representing different levels of computational complexity. As the batch size increases, the computational complexity increases as well.  We use 100 batches for every experiment as the input for all source nodes.

\textbf{Baselines and Performance Metric.} 
The baselines and performance metrics remain the same as in the Jetson-based tests. The three baselines are AR-MDI, MS-MDI, and Local. The performance is evaluated using the ``Average Inference Time''.

\subsubsection{Results}
\begin{figure}[htbp]
\centering
\centerline{\includegraphics[clip, trim=0cm 0cm 0cm 0cm, width=0.5\textwidth]{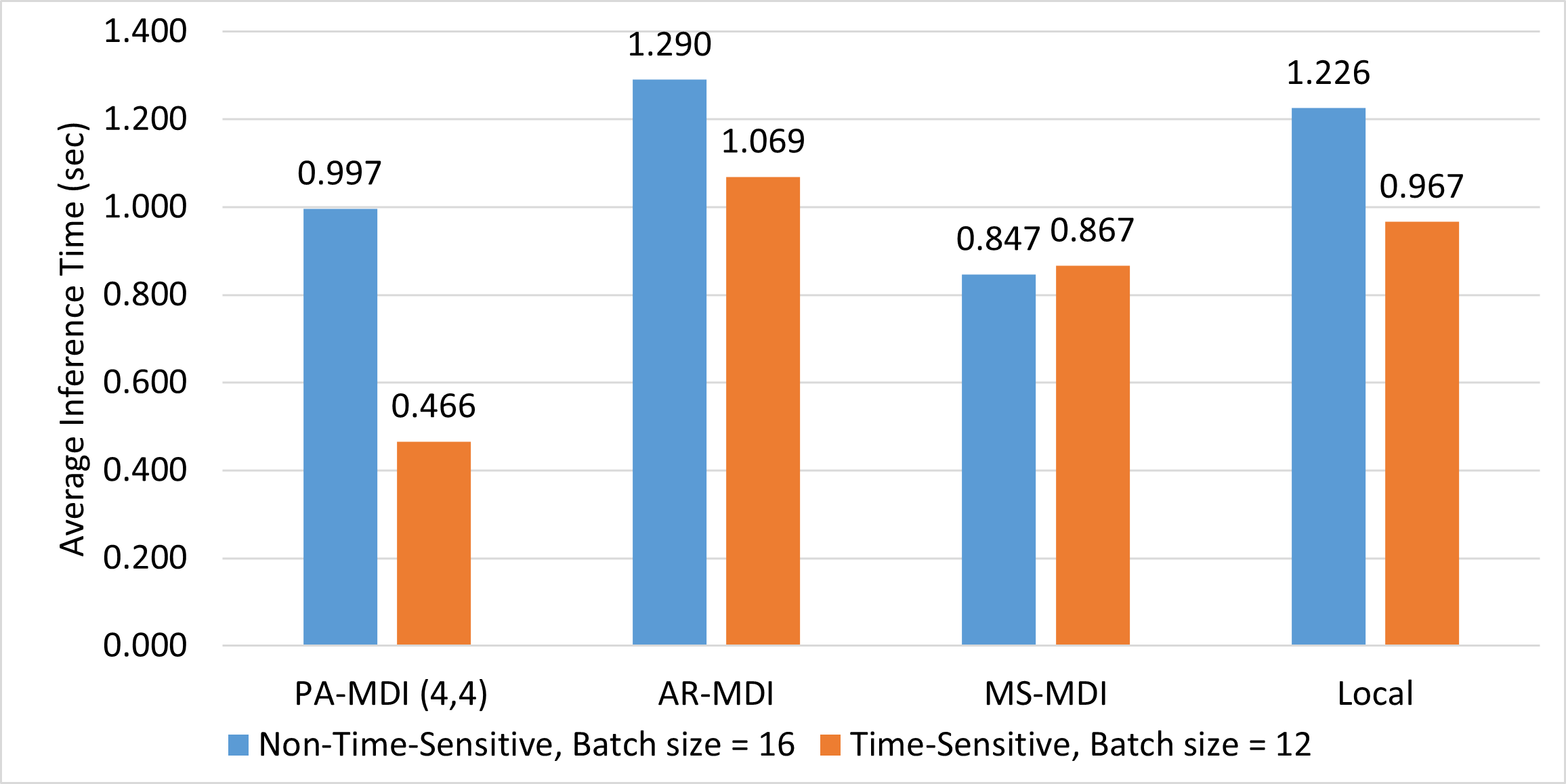}}
\caption{Workers $A$ has a larger input size (Batch Size = 16) and $D$ have smaller (Batch Size = 12).  Worker $A$ hosts ``Non-Time-Sensitive'' source, while Worker $D$ hosts ``Time-Sensitive'' source.
}
\label{fig:test4}
\vspace{-5pt}
\end{figure}

We evaluate the performance of \ours for two different scenarios. Similar to the previous tests, in the first scenario, Worker $A$, who hosts the ``Non-Time-Sensitive'' source, is assigned tasks with higher computational complexity by setting the batch size of tokens to 16. Meanwhile, Worker $D$, who hosts the ``Time-Sensitive'' source, is given tasks with lower computational complexity with a batch size of 12. The average inference time for \ours (4,4) compared to AR-MDI, MS-MDI, and Local is provided in Fig. \ref{fig:test4}.

The experimental results are similar to those observations in the first scenario of the Jetson-based test. However, in addition to \ours improving the average inference time compared to both AR-MDI and MS-MDI, \ours also demonstrated an advantage over the Local baseline. This is attributed to the fact that the communication cost in the LLMs-based experiments is significantly reduced due to the higher bandwidth in the Colosseum platform. Consequently, the benefits gained through MDI were more obvious. Compared to AR-MDI, MS-MDI, and Local, the average inference time for the ``Time-Sensitive'' source is reduced by up to 56.4\%, 34.8\%, and 51.8\%, respectively. 

\begin{figure}[htbp]
\centering
\centerline{\includegraphics[clip, trim=0cm 0cm 0cm 0cm, width=0.5\textwidth]{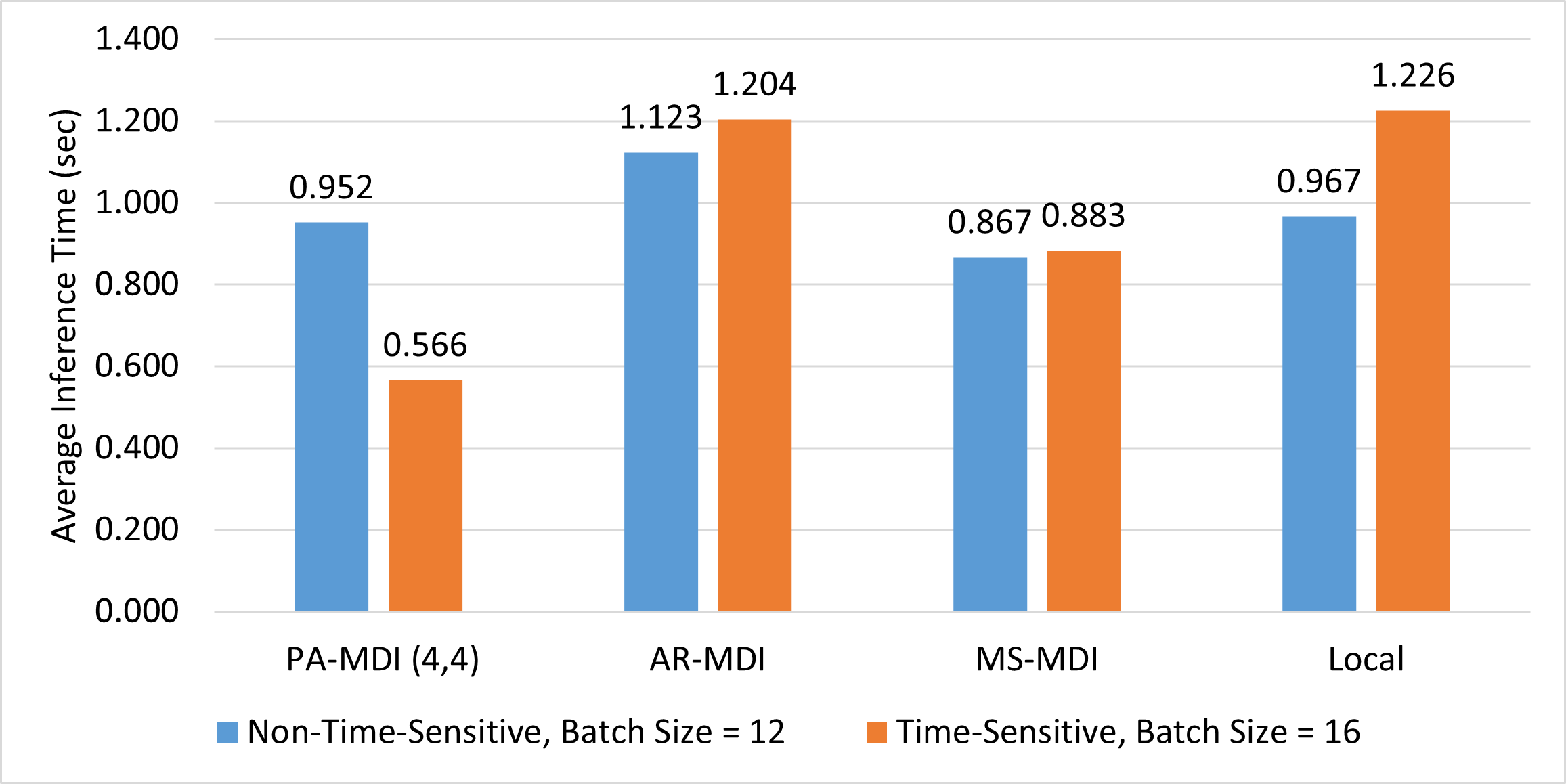}}
\caption{Worker $A$ has a smaller input size (Batch Size = 12) and $D$ have larger (Batch Size = 16). Worker $A$ hosts ``Non-Time-Sensitive'' data, while Worker $D$ hosts ``Time-Sensitive'' data.
}
\label{fig:test5}
\vspace{-5pt}
\end{figure}

In the second scenario, we reversed the computational complexity of the tasks. Worker $A$, responsible for the ``Non-Time-Sensitive'' source, is allocated tasks with a lower computational load by token batch size of 12.  Meanwhile, Worker $D$, managing the ``Time-Sensitive'' source, is given tasks with a higher computational load by a batch size of 16. The comparative average inference time for \ours (4,4) against AR-MDI, MS-MDI, and Local is illustrated in Fig. \ref{fig:test5}.

In the second scenario, \ours demonstrates a significant advantage over all baselines in the ``Time-Sensitive'' data. Compared to AR-MDI, MS-MDI, and Local, the average inference time for the ``Time-Sensitive'' data is reduced by up to 53.0\%, 35.9\%, and 53.9\%, respectively.

Overall, in our LLM testing, 
\ours significantly reduces the average inference time compared to both AR-MDI and the Local baselines, for both ``Time-Sensitive'' and ``Non-Time-Sensitive'' data. Additionally, \ours showed a performance advantage over MS-MDI in the ``Time-Sensitive'' data, with the cost that a mild overhead increase in the lower-priority ``Non-Time-Sensitive'' data.

%% file: conclusion.tex
\section{Conclusion}
\label{sec:conc}

We investigated the design and development of priority-aware model-distributed inference, where (i) ML models are partitioned and distributed across multiple workers, and (ii) we can prioritize ML models considering the importance of their data. We formulated an optimization problem to maximize ML inference accuracy while minimizing inference delay. Then, based on the structure of the optimization problem and its solution,  we designed a practical Priority-Aware Model-Distributed Inference (\ours) algorithm.  Experiments were conducted on a real-life testbed of NVIDIA Jetson Xavier edge devices as well as in the Colosseum testbed with ResNet-50, ResNet-56, and GPT-2 models. The experimental results showed that \ours performs priority-aware model allocation successfully while reducing the inference time as compared to baselines. 